# Nonlinear-Programming-Based Model of Power System Marginal States: Theoretical Substantiation

Boris I. Ayuev, Viktor V. Davydov, Petr M. Erokhin

*Abstract* - **In order to maintain the security of power system at an appropriate level and at low cost, it is essential to accurately assess the steady-state stability limits and power flow feasibility boundaries, i.e., the power system marginal states (MS). This paper is devoted to creation and theoretical substantiation of the MS model based on nonlinear programming (NLP-MS model), its research to reveal MS properties which promote better MS understanding, to evolution of the theory of power systems and MS, to elaboration of more effective algorithms of MS problem solution. The proposed NLP-MS model is universal and allows to determine and to take into account various MS, including the MS in a given direction of power change, the closest MS, a moving the power system state into a power flow feasibility region, etc.**

*Index Terms*— **Eigenvector, Jacobian, Lagrange multipliers, load flow, marginal state, nonlinear programming, power system, singularity, slack bus, stability, steady state.**

## I. INTRODUCTION

THE ASSESSMENT of marginal states (MS) plays the key role in analysis, planning and control of power systems. The MS is such stable steady state for which an arbitrarily small change in any of operating quantities in an unfavorable direction causes voltage collapse or loss of synchronism by units [1]. Usually a "try-and-error method" is used for searching MS, i.e., it is supposed that the initial steady state is stable, and then this state varies continuously until the determinant of the dynamic Jacobian becomes singular. With the introduction of the infinite bus as well as several other assumptions, monitoring the load flow Jacobian determinant can be used for detection of a possible aperiodic instability [2], [3]. In this case the load flow Jacobian determinant for the marginal aperiodic stable steady state is equal to zero. Therefore one of the most commonly used methods consists of sequential load flow solutions with the given step along the specified trajectory until a divergence of load flow solution is obtained, which in its turn is corrected by the binary search [4] or by the damped Newton method [2],

[5]. However, as soon as the network comes close to the condition of instability, a divergence in load flow solution may occur that is caused by the numerical problems of ill-conditioned Jacobian. That is why, in continuation methods the load flow parameterization [6]-[8] or the normalized iteration step change [9], [10] are widespread. In this case the Jacobian of intermediate states and marginal state is not singular, which allows to get "nose" curves [11].

The MS equations [12], which are more known as the point of the collapse method (PoC) [13]-[14] and which contain a set of nodal load flow equations and equations of linear dependence of columns or rows of Jacobian, play a significant role in MS studies. Generalized MS equations (GMS) in [15]-[16] allow to determine different MS including the ones which are the closest to the initial steady state. In [17] the closest MS equations based on geometric interpretation of the left eigenvector of singularity Jacobian as a normal vector to the MS hypersurface are given for Euclidean metric. The PoC and GMS include a double amount of variables and equations in comparison with the standard load flow problem but in case of success they significantly surpass the continuation methods [14]. However, there is a significant obstacle: good initial guesses for the system variables are essential; otherwise the Newton approach for obtaining the solution to the PoC (GMS) equations either yields undesirable results or does not converge.

A rather effective and universal approach is to use the optimization models and techniques to study MS problems. It appears that many MS problems can be represented as optimization problems. It allows to consider various constraints, to expand a range of solved MS problems, to use the powerful tools of nonlinear programming. The first publications appeared almost a quarter of a century ago. In [15] - [16] ties between optimization techniques and MS problems have been shown and the approach for finding various MS, including the closest MS, by means of optimization tools was proposed. In [18]-[20] MS problems in a given direction of power change are formulated as optimization problems. In [21], the interior point optimization technique is used to move a system state into power flow feasibility region. In [17], [22], a search for the closest MS is formulated as an optimization problem. In [23], the gradient method is used for re-dispatching a generation to increase a power system security margin. In [24] - [31] optimization techniques are used to provide a necessary security margin or to solve other MS problems.

The main feature of the optimization techniques proposed to determine MS or to incorporate MS in other power system

---

B. I. Ayuev is with Open Joint-stock Company System Operator of the United Power System, Moscow, Russia.

V. V. Davydov is with Open Joint-stock Company System Operator of the United Power System branch – ODU of Siberia and also with East Siberia State University of Technology and Management, Ulan-Ude, Russia (email davv@bur.so-ups.ru)

P. M. Erokhin is with Open Joint-stock Company System Operator of the United Power System and also with Ural Federal University, Yekaterinburg, Russia (email epm@ural.so-ups.ru; ).

problems is that they are restricted by consideration of a MS in a given direction of power change, using the model proposed in [18]. To solve other MS problems, e.g., to search the closest MS, more complex optimization models [22], [23] are proposed. These models require considerable computation efforts; therefore they do not have wide application.

The objective of this paper is to propose, to theoretically substantiate and to analyze the MS model based on nonlinear programming (NLP-MS model) The proposed model is universal. A number of existing optimization-based models of MS, e.g., presented in [15], [16], [18]-[21], [25], [31]-[34], are special cases of this model. The NLP-MS model allows to determine and to take into account various MS, including the MS in a given direction of power change, the closest MS, moving the power system state into a power flow feasibility region etc., to raise reliability and accuracy of computations. Moreover, the NLP-MS model has an important theoretical meaning. It allows to reveal important new properties of MS, which promote better understanding of power flow and MS models, development of the MS theory and elaboration of more effective algorithms for the solution of MS problems.

The rest of the paper is organized as follows. In Section II an analysis of peculiar properties of the load flow model and MS is performed. On the basis of this analysis in Section III the NLP-MS model is proposed, the NLP-MS models with a single and a distributed slack bus are studied, the important properties of MS are revealed, and numerical examples are presented. In Section IV the main results and conclusions are summarized, and the consideration subjects of following papers are also presented.

## II. Peculiar Properties of Load Flow Model

The load flow solutions are so common in the power systems that terminology and concepts of the load flow model are widely used for solving other power system problems. For example, the slack bus concept has been used to develop the effective optimization algorithms and solve the optimal power flow problems [4]. It has been found out that the choice of the slack bus location does not influence the optimal power flow (OPF); however, it influences the convergence of the iterative process. Therefore in these algorithms the slack bus is selected mainly from the point of view of convergence. Similar approach was used in earlier load flow algorithms, where selecting the slack bus considerably influenced their convergence. As a rule, "the slack bus is now considered a mathematical artifact created by load flow analysts, without any direct link with the physical system. Usually, the largest generator is arbitrarily proposed as slack bus in the absence of better criteria" [35]. Like in the case of the standard load flow solutions, in the majority of works on MS, including using optimization techniques, it is implicitly assumed that the slack bus location makes small impact on the ultimate solutions [36]. It is true for the model of lossless power systems, but it is absolutely incorrect for the realistic power systems MS [31], [32]. The slack bus location determines MS to a great extent. In MS, a change of slack bus position makes the steady state non-marginal [32]. In this case the power system steady state remains the same but the saddle-node bifurcation, the point of system collapse or the point of voltage collapse "vanish" at once. The calculated MS also depend upon other components of the load flow model. Therefore to create the NLP-MS model it is necessary to study the features of the standard load flow model and MS.

### A. Analysis of Load Flow Equations

Consider the nodal load flow equations in the polar coordinate system

$$\Delta P_k = P_k + \sum_m V_k V_m |Y_{km}| \sin(\delta_{km} - \alpha_{km}) = 0,$$
$$\Delta Q_k = Q_k - \sum_m V_k V_m |Y_{km}| \cos(\delta_{km} - \alpha_{km}) = 0, \quad (1)$$

where $\delta_k$, $V_k$ are the voltage angle and magnitude at bus $k$; $P_k$, $Q_k$ are active and reactive powers at bus $k$; $Y_{km}$ is a component of the bus admittance matrix; $\alpha_{km} = \angle Y_{km} + \pi/2$ is a loss angle; $\delta_{km} = \delta_k - \delta_m$.

The system of the nonlinear equations (1) may be rewritten in the compact form:

$$\Delta F(X,Y) = 0, \quad (2)$$

where $\Delta F(X,Y)$, $X$, $Y$, are the vector of power mismatches and the vectors of dependent and independent variables respectively. Partitioning the variables into dependent and independent ones reflects the fact that the number of variables (parameters) in (2) is bigger than the number of equations. Mathematically there is no special way to determine the dependent and independent variables. Only the number of the dependent variables is given. Each known variable is set as independent. From the point of view of the load flow solutions the independent variables are those that can be controlled [37], e.g., power injections and voltage magnitudes at the generation buses with automatic excitation control. Bus voltages are dependent variables resulting from the solution of (2).

A solution of the nonlinear equations system (2) may be considered as a mapping of independent variables into the dependent variables space $X=X(Y)$ [38]. According to the implicit function theorem the essential prerequisite condition of the mapping existence (and consequently the solution in general) is non-singularity of the matrix of partial derivatives of the system of nonlinear equations with respect to dependent variables $[\partial \Delta F/\partial X]$.

Consider the full matrix of partial derivatives (Jacobian) of the nodal load flow equations with respect to the voltage angles and magnitudes:

$$[J_F] = \begin{vmatrix} \partial \Delta P/\partial \delta & \partial \Delta P/\partial V \\ \partial \Delta Q/\partial \delta & \partial \Delta Q/\partial V \end{vmatrix}, \quad (3)$$

where

$$\left(\frac{\partial \Delta P}{\partial \delta}\right)_{km} = \begin{cases} -V_k V_m |Y_{km}| \cos(\delta_{km} - \alpha_{km}), \ k \neq m; \\ \sum_{m \neq k} V_k V_m |Y_{km}| \cos(\delta_{km} - \alpha_{km}), \ k = m; \end{cases}$$

$$\left(\frac{\partial \Delta P}{\partial V}\right)_{km} = \begin{cases} V_k |Y_{km}| \sin(\delta_{km} - \alpha_{km}), \ k \neq m; \\ 2V_k ReY_{kk} + \sum_{m \neq k} V_m |Y_{km}| \sin(\delta_{km} - \alpha_{km}), \ k = m; \end{cases}$$

$$\left(\frac{\partial \Delta Q}{\partial \delta}\right)_{km} = \begin{cases} -V_k V_m |Y_{km}| \sin(\delta_{km} - \alpha_{km}), \ k \neq m; \\ \sum_{m \neq k} V_k V_m |Y_{km}| \sin(\delta_{km} - \alpha_{km}), \ k = m; \end{cases} \quad (3a)$$

$$\left(\frac{\partial \Delta Q}{\partial V}\right)_{km} = \begin{cases} -V_k |Y_{km}| \cos(\delta_{km} - \alpha_{km}), & k \neq m; \\ 2V_k \, Im \, Y_{kk} - \sum_{m \neq k} V_m |Y_{km}| \cos(\delta_{km} - \alpha_{km}), & k = m. \end{cases}$$

Equations (3) and (3a) show that the diagonal components of $[\partial \Delta P/\partial \delta]$ and $[\partial \Delta Q/\partial \delta]$ are equal to the sum of the non-diagonal components in the row with the opposite sign. Therefore $[\partial \Delta P/\partial \delta]e=0$, $[\partial \Delta Q/\partial \delta]e=0$, where $e$ is a vector with all components equal to one. Hence the full Jacobian $[J_F]$ will be singular, since $[J_F][e^T, 0^T]^T = [0^T, 0^T]^T$.

This feature of the full Jacobian (3) is a direct consequence of (1). Shifting of every phase angle on the same value does not change the left side. Therefore, the system (1) has infinite number of solutions. It is necessary to set a reference point. In order to do it the voltage angle of a bus should be considered as known, i.e., this voltage angle is moved from dependent to independent variables. The bus is called the angle reference bus. Since the number of dependent variables decreases by one it is necessary to restore it. Due to the power system features the active power of a bus becomes a new dependent variable. Such bus becomes a balancing active power bus and is named the slack bus. The reference bus may be chosen arbitrarily. The slack bus choice is determined by the power system specifics.

The necessity to use a slack bus for load flow solutions has an interesting geometrical interpretation. In [39] it is shown that a projection of the load flow solution space onto the space of active powers of the buses and reactive powers of the PQ buses is a hypersurface in this bus powers space. The authors called such hypersurface the hypersurface of bus powers of the power system steady states (HBPS). Each power system steady state corresponds to a certain point on the HBPS and vice versa. This HBPS is a mapping of the whole set of power system steady states, including all hypothetical (unstable) power system steady states, into this space of bus powers. As any hypersurface, the HBPS has a (Lebesgue) measure[1] zero, on the bus powers space [40]. In terms of probability theory it means that a chance to specify the coordinates of a point on HBPS (the bus powers) without taking into account their functional dependence so that they could satisfy the equation(s) of hypersurface (load flow), is equal to zero. Hence, to obtain the coordinates of a point on the HBPS (to solve the load flow equations) it is necessary to use a coordinate of this point (active power of a slack bus) as a dependent variable. Therefore all computational models that use the load flow equations always apply a slack bus explicitly or implicitly – not only load flow solutions, but also the optimal power flow, estimation of the power system state, a steady-state stability assessment, etc. [41] - [42].

In general, two other sub-matrices $[\partial \Delta P/\partial V]$ and $[\partial \Delta Q/\partial V]$ of the full Jacobian are not singular. However, if there are no shunts during a flat start, their diagonal components are also equal to the sum of non-diagonal components in the row with the opposite sign, i.e., $[\partial \Delta P/\partial V]e=0$, $[\partial \Delta Q/\partial V]e=0$. In this case $[J_F][0^T, e^T]^T = [0^T, 0^T]^T$ and the dimension of the null space of the full Jacobian will be equal to two. The use of the reference and slack buses reduces the dimension of the null space of the full Jacobian by one. However, the Jacobian remains singular. In most cases voltage magnitudes are not equal to the flat start values, therefore sub-matrices $[\partial \Delta P/\partial V]$ and $[\partial \Delta Q/\partial V]$ are not singular. They are, however, ill-conditioned and the resulting solution is very sensitive to reactive power variation. Consequently, it is practically impossible to set the bus reactive powers in such a way, that the resulting state could correspond to the operating conditions [37]. Therefore it is necessary to fix the voltage magnitude for one of the buses and to make it an independent variable. As it was in the case of voltage angle setting, the reactive power of one of the buses becomes a dependent variable. This bus becomes the balancing reactive power bus. Usually, the balancing active power bus (slack bus) does the same with the reactive power. It has a given voltage amplitude and angle ($V\delta$-bus). Generator buses equipped with automated excitation control balance the reactive power as well. These buses keep voltage magnitudes unchanged (PV-buses).

Taking into account all the aforementioned, the matrix of partial derivatives of load flow equations with respect to dependent variables turns into

$$[J] = \begin{bmatrix} J_{LF} & 0 & 0 \\ \partial \Delta P_b/\partial \delta & \partial \Delta P_b/\partial V_{PQ} & 1 & 0 \\ \partial \Delta Q_{PV}/\partial \delta & \partial \Delta Q_{PV}/\partial V_{PQ} & 0 & E \end{bmatrix}, \quad (4)$$

where

$$[J_{LF}] = \begin{bmatrix} \partial \Delta P/\partial \delta & \partial \Delta P/\partial V_{PQ} \\ \partial \Delta Q_{PQ}/\partial \delta & \partial \Delta Q_{PQ}/\partial V_{PQ} \end{bmatrix} \quad (5)$$

is the standard load flow Jacobian, $[E]$ и $[0]$ are the identity and zero sub-matrices of respective dimensions, and index $b$ is used for the slack bus. It follows from (4) that $det[J]=det[J_{LF}]$ and non-singularity of the load flow Jacobian (5) ensures non-singularity of matrix (4).

Formally, to solve the system of nonlinear equations (1) by the Newton method, the system of linearized equations with matrix (4) is to be solved. However, the linearized equations using $[J_{LF}]$ are independent from other equations and are solved separately. All the other dependent variables, e.g., the slack bus power and reactive powers at PV-buses are determined from (1) by simple substitution of voltage angles and amplitudes.

Several load flow programs use the so-called distributed slack bus, i.e., the active power balancing is provided by several generators with given participation factors. Participation factors $\alpha_k$ may be given according to economic reasons, or based on the requirements of primary or secondary frequency control. In this case dependent variable $P_S$ is used for the distributed slack bus. It is considered in the load flow equations as $\alpha_k P_S$ with $\sum \alpha_k = 1$, and the Jacobian is represented as follows

$$[J_{LF}^S] = \begin{bmatrix} \partial \Delta P/\partial \delta & \partial \Delta P/\partial V_{PQ} & \alpha \\ \partial \Delta Q_{PQ}/\partial \delta & \Delta Q_{PQ}/\partial V_{PQ} & 0 \end{bmatrix}, \quad (6)$$

where $\alpha$ is the vector of bus participation factors for active power balancing. The matrix (6) includes partial derivatives of equations of the active powers balance for all buses.

---

[1] In mathematical analysis, the Lebesgue measure is the standard way to assign a measure to subsets of n-dimensional Euclidean space. For n=1, 2, or 3, it coincides with the standard measure of length, area, or volume. In general, it is also called n-dimensional volume, n-volume, or simply volume.

## B. MS Peculiarities

The active and reactive powers at buses are included in (1) additively. Excluding technical limitations, the system (1) will always be consistent with any voltage angles and magnitudes, if bus powers are derived from (1) directly. Therefore MS is a steady state where a small deviation of independent variables leads to inconsistency of the system (1), i.e., it may not be solved concerning the dependent variables. The implicit function theorem asserts that if the Jacobian matrix (4) at point $X_0$, $Y_0$ is not singular, then for each $Y$, close enough to $Y_0$, there is only one solution $X=X(Y)$, which in its turn is the solution of nonlinear nodal load flow equations $\Delta F(X(Y),Y)=0$. Therefore the implicit function theorem corollaries are: firstly, a necessary condition (criterion) for the MS is the singularity of (4)-(6) [12], secondly, another necessary condition is the existence of close alternative solution around the MS [40]. According to (4) - (6), Jacobian depends upon the set of dependent and independent variables of the load flow model. Therefore a slack bus location influences the load flow Jacobian and MS.

It is necessary to mention an important property of the load flow Jacobian. According to the Sard theorem, a set of non-regular values of variables, i.e., when the Jacobian is singular, has measure zero [38]. In terms of the probability theory it means that the probability of such events is equal to zero. The complement for every set of measure zero is dense everywhere, i.e., the set of regular values has full measure. Therefore each point in the variables space is arbitrarily close to some regular value of mapping. On the one hand, the Sard theorem guarantees the impossibility of practical existence of normal steady state having singular Jacobian, i.e., it makes the necessary condition of MS sufficient. On the other hand, it asserts that the obtained solutions are only approximations of actual MS. The Sard theorem is seldom applied by mathematicians. However, it is very powerful and useful mathematical tool for those who elaborate or apply computing models of power systems. E.g., the theorem allows to explain the operability of the damped Newton method when the load flow equations have no real solution. It is known that in this case the iteration process of the damped Newton method comes to an MS with a "singular" Jacobian [29], [37]. Theoretically, it is impossible to solve a system of linear equations with the singular matrix. However, it is done by the damped Newton method [30]. It confirms that in computing models the load flow Jacobian is not singular though it may be ill-conditioned. In the damped Newton method an optimal multiplier considerably improves the condition number of the solved problem.

The main feature of the existing ways to search the MS is that an MS criterion is added to the computational model of the load flow as a "forced tool" to "push" a system state to an MS. The MS may be compared to a "bluff". When the system state is far from MS, the computational models almost do not feel the edge of the "bluff" and may "fly" far beyond the limits [12]. Therefore it is not surprising that commercial programs compute a MS using a preset direction with a discrete or parameterized step until iteration fails. At the same time the MS criterion is implicitly included into the computational model of the power system steady states [15].

## III. NLP-MS MODEL AND ITS ANALYSIS

### A. NLP- MS Model

Singularity of the load flow Jacobian, as well as an MS, depends upon a set of dependent and independent variables. First of all, the MS is a steady state, but some of its parameters have marginal (extreme) values. Using a variables set as in the load flow models (1)-(6), consider the nonlinear programming problem with equality constraints.

$$\min f(X,Y) \qquad (7)$$
$$\text{s.t.} \quad \Delta F(X,Y) = 0 . \qquad (8)$$

Here (7) is an objective function, and (8) is load flow equations in the form of (2). The vector of optimized variables consists of two vectors $X$ and $Y$. The nonlinear programming theory considers all variables as equal in rights and does not distinguish between them. The partition has been introduced while developing the reduced gradient method [4] for OPF. Being based on the implicit function theorem the optimization variables are grouped into dependent $X$ and independent $Y$ variables. That allows to significantly decrease the complexity of OPF solution, decoupling the problem into two parts: a load flow solution (solving (8)) and an optimization step in the space of independent variables $Y$.

The Lagrange function for (7) - (8) can be represented as

$$L(X,Y,\lambda) = f(X,Y) + \Delta F(X,Y)^T \lambda . \qquad (9)$$

Here $\lambda$ is a vector of auxiliary solution variables, called Lagrange multipliers.

The first-order necessary optimality conditions for the problem (7)-(8) are given by the following equations

$$\nabla_X L = \nabla_X f + [\partial \Delta F / \partial X]^T \lambda = 0 ; \qquad (10)$$
$$\nabla_Y L = \nabla_Y f + [\partial \Delta F / \partial Y]^T \lambda = 0 ; \qquad (10a)$$
$$\nabla_\lambda L = \Delta F(X,Y) = 0 , \qquad (10b)$$

where $\nabla_X L = [\partial L / \partial X]^T$ is a gradient of function $L$ with respect to vector $X$.

If the gradient of objective function (7) with respect to the dependent variables is equal to zero, i.e., $\nabla_X f = 0$, then the condition (10) will be transformed into

$$\nabla_X L = [\partial \Delta F / \partial X]^T \lambda = 0 . \qquad (11)$$

In its turn, if vector $X$ of the nonlinear programming problem (7) - (8) corresponds to the vector of dependent variables of the load flow problem then the matrix $[\partial \Delta F / \partial X]$ in (11) will correspond to the Jacobian (4). In this case (11) with $\lambda \neq 0$ determines singularity of Jacobian (4), i.e., MS.

In order to guarantee (11), it is enough to exclude the dependent variables from the objective function (7). In this case the solution of the nonlinear programming problem (7) - (8) will be MS corresponding to a minimum of the objective function (7). By means of this model called the NLP-MS model, it is possible to determine various MS, e.g., the closest MS. For this purpose it is necessary to use the corresponding objective function. Applications of the NLP-MS model for MS problems solution will be presented in the following papers.

The NLP-MS model also has an important theoretical meaning. Its analysis allows to reveal important properties of MS which promote better understanding of MS and power flow models, and development of the MS theory.

## B. Analysis of NLP- MS Model With Single Slack Bus

Using the properties of the NLP-MS model, consider OPF model

$$\min f(P,Q) \qquad (12)$$

s.t.

$$\begin{bmatrix} \Delta P(P,\delta,V) \\ \Delta P_b(P_b,\delta,V) \\ \Delta Q(Q,\delta,V) \end{bmatrix} = \begin{bmatrix} 0 \\ 0 \\ 0 \end{bmatrix} \qquad (13)$$

The system (13) is a vector representation of the load flow equations (1). The objective function is determined by the problem being solved. All variables in the optimization problem (12)-(13) are equal in rights. However, due to singularity of the full Jacobian, one of the buses should have a fixed voltage angle and at least one bus should have a fixed voltage magnitude. There are no other constraints. To associate the solution of (12)-(13) with an MS, it is necessary to use the set of dependent and independent variables similar to the load flow problem. $b$ represents a slack bus. It can be any bus participating in the optimization problem.

The Lagrange function for (12) - (13) can be represented as

$$L = f - \Delta P^T \lambda^P - \Delta P_b \lambda_b^P - \Delta Q^T \lambda^Q. \qquad (14)$$

The first-order optimality conditions for the problem (12)-(13) are given by the following equations

$$\partial L / \partial P_m = \partial f / \partial P_m - \lambda_m^P = 0; \qquad (15)$$

$$\partial L / \partial P_b = \partial f / \partial P_b - \lambda_b^P = 0; \qquad (15a)$$

$$\partial L / \partial Q_m = \partial f / \partial Q_m - \lambda_m^Q = 0; \qquad (15b)$$

$$\begin{bmatrix} \nabla_\delta L \\ \nabla_V L \end{bmatrix} = -\begin{bmatrix} \partial \Delta P/\partial \delta & \partial \Delta P/\partial V \\ \partial \Delta P_b/\partial \delta & \partial \Delta P_b/\partial V \\ \partial \Delta Q/\partial \delta & \partial \Delta Q/\partial V \end{bmatrix}^T \begin{bmatrix} \lambda^P \\ \lambda_b^P \\ \lambda^Q \end{bmatrix} = 0. \qquad (15c)$$

Analysis of the linearized equations (15c) allows to reveal important properties of Lagrange multipliers used in the NLP-MS model.

The equations system matrix (15c) is rectangular. It includes the load flow Jacobian (5) plus the row of partial derivatives of the slack bus active power equation. Add the column of partial derivatives of load flow equations with respect to the voltage angle of the slack bus and obtain the augmented Jacobian

$$[J_A] = \begin{bmatrix} J_{LF} & & \partial \Delta P/\partial \delta_b \\ & & \partial \Delta Q_{PQ}/\partial \delta_b \\ \partial \Delta P_b/\partial \delta & \partial \Delta P_b/\partial V_{PQ} & \partial \Delta P_b/\partial \delta_b \end{bmatrix}. \qquad (16)$$

The augmented Jacobian as well as the full Jacobian (3) is singular. Its right eigenvector corresponding to zero eigenvalue is equal to $[e^T, 0^T, 1]^T$. Its left eigenvector corresponding to zero eigenvalue is the vector of Lagrange multipliers $\lambda = \begin{bmatrix} \lambda^{P^T}, \lambda^{Q^T}, \lambda_b^P \end{bmatrix}^T$ of the optimization problem (12)-(15c). To see it, it is enough to check up the following expression:

$$[\partial \Delta P/\partial \delta_b]\lambda^P + \partial \Delta P_b/\partial \delta_b \lambda_b^P + [\partial \Delta Q/\partial \delta_b]\lambda^Q = 0. \qquad (17)$$

Since the right eigenvector of the augmented Jacobian, corresponding to zero eigenvalue is equal to $[e^T, 0^T, 1]^T$, the multipliers $\partial/\partial \delta_b$ in (17) may be determined through summation of the row elements of the corresponding sub-matrix. Therefore taking into account (15c) yields (17).

The augmented Jacobian (16) also allows to obtain a geometrical interpretation of the vector of Lagrange multipliers. The using of its properties in the linearized load flow equations results in the vector of Lagrange multipliers $\lambda$ as a normal vector to the HBPS [39].

The system (15c) also determines the interrelation between Lagrange multipliers. If to rewrite (15c) as

$$\begin{bmatrix} \partial \Delta P/\partial \delta & \partial \Delta P/\partial V \\ \partial \Delta Q/\partial \delta & \partial \Delta Q/\partial V \end{bmatrix}^T \begin{bmatrix} \lambda^P \\ \lambda^Q \end{bmatrix} = -\begin{bmatrix} \partial \Delta P_b/\partial \delta^T \\ \partial \Delta P_b/\partial V^T \end{bmatrix} \lambda_b^P \qquad (18)$$

and to take advantage of the reduced gradient approach, the following expressions are obtained (see Appendix):

$$\lambda_m^P = (1 - \partial \pi/\partial P_m)\lambda_b^P, \qquad (19)$$

$$\lambda_m^Q = -\partial \pi/\partial Q_m \, \lambda_b^P, \qquad (19a)$$

where $\partial \pi/\partial P_m$ is an incremental transmission loss coefficient (ITL) for bus $m$.

The expressions (19)-(19a) together with (15)-(15b) give the power flow optimality criteria. The optimal solution does not depend upon a slack bus choice. The classical optimal power flow problem uses the slack bus as a reference point for power stations performance assessment. The difference between the right and left parts of (19) is a value of the reduced gradient vector element in the active power flow optimization problem. The optimality conditions (15)-(15b) and (19)-(19a) will remain unchanged in case of other slack bus assignment. Indeed, the Lagrange multipliers are the left eigenvector components of the augmented Jacobian (16) corresponding to zero eigenvalue. The augmented Jacobian remains the same regardless of a slack bus position. In fact the values of (1-$\partial \pi/\partial P_m$) will be different, but in (19)-(19a) they will change with the same ratio $1-\partial \pi_k/\partial P_m=(1-\partial \pi_b/\partial P_m)/(1-\partial \pi_b/\partial P_k)$, where $\partial \pi_b/\partial P_m$ is the ITL with the slack bus $b$, $\partial \pi_k/\partial P_m$ is the same with a new slack bus $k$.

Interrelations (15)-(15a) and (19) between Lagrange multipliers allow to reveal an interesting interpretation of MS if to consider the electric power trade at the wholesale market with the following objective function

$$f = \sum C_m P_m .$$

Here $C_m$ is a bid price of power plant $m$ for one kWh.

Consider a hydroelectric power station (HPS) $k$ with an overflowing pond. It is necessary either to dump the surplus, or to sell the electric power at any price. The HPS submits a very low price bid. Suppose also that due to repairing campaign the transmission capacity of the station transmission lines is limited. How much energy will the HPS sell? The answer is in (15) and (19) if considered as follows $\lambda_k^P=C_k=(1-\partial \pi/\partial P_k)C_b$. Therefore, if $C_k<<C_b$, then $1-\partial \pi/\partial P_k= C_k/C_b \approx 0$. When the HPS announces a price-taking behavior, $\lambda_k^P=C_k=0$ and $\partial \pi/\partial P_k=1$, i.e., the station will be loaded in such a way that any additional increase in generation will be compensated by losses associated with such increase. If this station is not taken as the slack bus, then the steady state will not be an MS, since its additional loading is possible. As soon as its ITL will exceed one, the additional increase of generation will cause decreasing of effective output. It will be compensated by a slack bus. If this station is taken as the slack bus for the load flow solution, then the steady state will be marginal. In this case (15a) and (18) will be transformed to the following

$$\lambda_b^P = 0 \qquad (20)$$

$$\begin{vmatrix} \partial \Delta P/\partial \delta & \partial \Delta P/\partial V \\ \partial \Delta Q/\partial \delta & \partial \Delta Q/\partial V \end{vmatrix}^T \begin{bmatrix} \lambda^P \\ \lambda^Q \end{bmatrix} = \begin{bmatrix} 0 \\ 0 \end{bmatrix}. \qquad (20a)$$

The matrix in (20a) is the load flow Jacobian (5) and the condition (20a) together with $\lambda^P$, $\lambda^Q \neq 0$ determines its singularity, therefore it is an MS.

Thus, an MS is inability of the slack bus to maintain the steady state. Its every movement towards balancing an active power is leveled by losses associated with such balancing inputs. The slack bus as if loses the connection with a part or the whole power system; it corresponds to absence of the slack bus in the computational model with all consequences.

Zero price bid means exclusion from the objective function. Therefore, if the active power of the slack bus is excluded from (12), then the solution of (12)-(13) will be a MS corresponding to the minimum of the objective function.

The singularity of the load flow Jacobian (20a) does not break the conditions of the Lagrange function use and the existence of an implicit function in the MS. According to the nonlinear programming theory [4], in order to guarantee the existence and uniqueness of the Lagrange multipliers vector, the gradients of the constraint equations (13) should be linearly independent. In the optimization problem (12)-(13) the number of variables exceeds the number of constraint equations. Therefore, the homogeneous system of linear dependence of gradients has more equations than variables. According to the Sard theorem, a non-zero solution of such a system, i.e. a linear dependence of gradients, has measure zero [40]. The linearly independent set of the equality constraints gradients (13) ensures existence conditions of an implicit function in the MS for every bus having a non-zero Lagrange multiplier in (20a), but not for the slack bus. This feature is used in the continuation methods [7], [9]-[11] for changing the set of dependent and independent variables. In the MS the dimension of the null space of the augmented Jacobian (16) will be equal to one [41], as in any other steady state. The left eigenvector for zero eigenvalue of the augmented Jacobian will also correspond to the vector of Lagrange multipliers (20) - (20a). This may be shown by examining (16).

In the MS, in case of changing the slack bus, the conditions (20)-(20a) will not be satisfied generally, unless a new slack bus has the zero Lagrange multiplier in (20a). A simple system with 4 *PV* buses in Fig. 1 allows to show it.

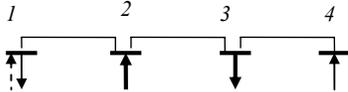

Fig. 1. 4-bus system

The bus powers in MW (the minus sign corresponds to a generation) and Lagrange multipliers for the base case and also for the MS with the slack bus 4 as well as the slack bus 1 are shown in Table I. These MS are the results of generation increase at bus 2 and load increase at bus 3. The line parameters are $Z_{12}=Z_{34}=5+j10$ $\Omega$, $Z_{23}=20+j40$ $\Omega$ and $V_1=V_2=V_3=V_4=110$ $kV$.

TABLE I
4-BUS SYSTEM

| Bus | 1 | 2 | 3 | 4 |
|---|---|---|---|---|
| Base Case | | | | |
| P | 20 | -50 | 50 | -22.1951 |
| λ | 0.9259 | 0.9069 | 1.0231 | 1 |
| MS With Slack Bus 4 | | | | |
| P | 20 | -411.77 | 411.77 | -465.27 |
| λ | 0.7144 | 0.6997 | 0 | 0 |
| MS With Slack Bus 1 | | | | |
| P | -118.72 | -171.51 | 171.51 | -22.1951 |
| λ | 0 | 0 | 0.7151 | 0.6990 |

Consider the MS parameters with the slack bus 4. Since power transfer through line 2-3 is associated with losses, the slack bus 4 supports the power transfer through line 3-4. The MS takes place when line 2-3 is overloaded. The steady state remains marginal if the slack bus will be moved to bus 3 since its Lagrange multiplier is also equal to zero in this case. However it will not remain marginal, if the slack bus is moved to bus l as its Lagrange multiplier is not equal to zero and there is a possibility to continue generation increase at bus 2 and load increase at bus 3. In the MS the ITL for bus 4 with the slack bus 1 or 2 is equal to one.

MS will be different with slack bus 1 because it supports the system steady state through line 1-2 in this case. The ITL for the slack bus increases with the growth of power transfer. As soon as it reaches one, a load increase at bus 3 becomes impossible. Slack bus 1 is unable to support the steady state of bus 3 and 4 in this MS. The MS remains marginal if bus 2 becomes the slack bus, as its Lagrange multiplier is equal to zero. This steady state will not be marginal if the slack bus is bus 4 or 3 since these buses have non-zero Lagrange multipliers. The marginal values of bus powers and losses for system with slack bus 1 are smaller than with slack bus 4. At the same time the MS with slack bus 4 and the steady states around the MS will be unstable if the dynamic Jacobian is used for the steady-state stability assessment [42].

The influence of the location of slack bus on the MS can be explained by means of the HBPS. Each power system steady state corresponds to a certain point on the HBPS and vice versa. A projection of the HBPS along the axis direction of slack bus active power onto the subspace of all specified bus powers is the power flow feasibility region [39]. The boundary of this projection corresponds to MS set. As the HBPS is usually not "flat", the points on the HBPS which correspond to the boundary of such projection, i.e., MS, will be different for another slack bus. Therefore a slack bus change in MS makes the steady state non-marginal.

### C. Analysis of NLP-MS Model With Distributed Slack Bus

In case of a distributed slack bus its power $P_S$ is a dependent variable. The necessary optimality conditions (15) - (15c) remain unchanged but an additional equation is appeared

$$\partial f/\partial P_S - \alpha^T \lambda^P = 0. \qquad (21)$$

Interrelation between the Lagrange multipliers with the distributed slack bus is determined by the following expressions:

$$\lambda_m^P = (1 - \partial \pi_S/\partial P_m) \lambda^S; \qquad \lambda_m^Q = -\partial \pi_S/\partial Q_m \lambda^S, \qquad (22)$$

where $\partial \pi_S/\partial P_m$ is ITL for bus $m$ in the system with the distributed slack bus, and $\lambda^S=\alpha^T\lambda^P$.

If the variable $P_S$ is not included in the objective function (12), then (21) turns into

$$\sum \alpha_m \lambda_m^P = \lambda^S = 0 . \qquad (23)$$

Consideration of (15c) together with (23) yields singularity of the Jacobian (6), i.e., an MS.

A slack bus change in the MS does not change the Lagrange multipliers as they correspond to (15) - (15c). According to (22) the following expression is also true.

$$\lambda_m^P = (1 - \partial \pi_{Snew}/\partial P_m)\lambda^{Snew} , \qquad (24)$$

here, $\partial \pi_{Snew}/\partial P_m$ is ITL of bus $m$ in the system with the new distributed slack bus $s_{new}$. Therefore, if $\lambda^{Snew} \neq 0$, then a substitution of (24) into (23) yields

$$\partial \pi_{Snew}/\partial P_S = \sum (\partial \pi_{Snew}/\partial P_m)\alpha_m = 1 .$$

Thus, in the MS any manipulations of the slack bus to balance the active power are completely compensated by the active power losses caused by such manipulations. It means that the distributed slack bus is not able to support such steady state. Any random insignificant changes of powers at some buses cannot be balanced by the slack bus.

Similar to the case of a single slack bus, a change of the distributed slack bus (e.g., a change of the participation factors) in the MS makes this steady state non marginal. The exception is the case when the condition (23) will also be met in this steady state for a new structure of the distributed slack bus.

### D. Analysis of NLP-MS Model of Lossless Systems

In the case of no resistances in the network in Fig. 1 the MS of systems with slack bus 4 and with slack bus 1 become the same. The base case and MS parameters for 4-bus lossless system with $Z_{12}=Z_{34}=j10\ \Omega$, $Z_{23}=j40\ \Omega$ are shown in Table II

TABLE II
4-BUS LOSSLESS SYSTEM

| Bus | 1 | 2 | 3 | 4 |
|---|---|---|---|---|
| Base Case | | | | |
| P | 20 | -50 | 50 | -20 |
| λ | 1 | 1 | 1 | 1 |
| MS With Slack Bus 4 | | | | |
| P | 20 | -322.5 | 322.5 | -20 |
| λ | 1 | 1 | 0 | 0 |
| MS With Slack Bus 1 | | | | |
| P | 20 | -322.5 | 322.5 | -20 |
| λ | 0 | 0 | 1 | 1 |

In any lossless power system an MS remains marginal regardless to the slack bus location. It can be shown, if to consider the following. Firstly, if the resistances are equal to zero, then the left eigenvector of the augmented Jacobian (16) corresponding to zero eigenvalue, is always $[e^T, 0^T, 1]^T$. Secondly, in the MS the dimension of a null space of the augmented Jacobian (16) is equal to two [41]. Therefore the Lagrange multipliers in (20) - (20a) in the MS will correspond to other left eigenvector of the augmented Jacobian corresponding to zero eigenvalue. Hence, in the MS vector $\lambda$ of the linear combination of vectors

$$\lambda = \beta[e^T, 0^T, 1]^T + [\lambda^{P^T}, \lambda^{Q^T}, 0(=\lambda_b^P)]^T \qquad (25)$$

will also be an eigenvector of the augmented Jacobian corresponding to zero eigenvalue. Two cases are possible with a new slack bus $k$. If $\lambda^P_k=0$, then the Lagrange multipliers remain unchanged in (20) - (20a). Otherwise vector $\lambda$ (25) is a new vector of Lagrange multipliers with $\beta=-\lambda^P_k$. It also corresponds to the load flow Jacobian (5) singularity.

Similar situation appears with a change of the distributed slack bus in the MS for a lossless system. In this case vector $\lambda$ (25) with $\beta=-\sum\alpha_k\lambda^P_k$ correspond to (23), therefore the load flow Jacobian (6) is also singular.

The MS independence concerning the slack bus location can be also explained by means of the lossless power system HBPS. This HBPS is "flat" [39] since it also satisfies the equation $\sum P_k=0$. The points on "flat" HBPS, that correspond to the boundary of projection of this hypersurface along active power axis direction of any slack bus onto the subspace of the specified bus powers, i.e. the MS, are the same. Therefore the MS does not depend upon a slack bus location. It is also interesting to note that one "side" of this "flat" HBPS corresponds to aperiodic stable steady states, the other "side" is unstable. I.e., the same point on the lossless system HBPS corresponds to stable and unstable power system steady states.

### IV. CONCLUSION

In this paper the NLP-MS model of MS based on nonlinear programming is proposed, its theoretical substantiation and research are presented. The NLP-MS model allows to expand the range of solved MS problems and has an important theoretical meaning. It allows to reveal important MS properties which promote better understanding of load flow models and MS, the evolution of the MS theory.

The analysis of the NLP-MS model has shown that an MS depends upon a given set of dependent and independent variables of the load flow model and, especially, on a slack bus location. In the MS the incremental transmission loss coefficient of the slack bus is equal to one and so any attempts of the slack bus to support the power system steady state are completely compensated by the associated power loss. Inability of the slack bus to support a steady state even of some buses determines the power system MS. In MS the slack bus as if loses the connection with a part or the whole power system; it corresponds to absence of the slack bus in the computational model with all consequences. In realistic power systems a change of slack bus location in the MS makes this steady state non marginal. Only in lossless systems the MS do not depend upon a slack bus location.

The proposed NLP-MS model is universal. It allows to solve various problems associated with MS. Applications of the NLP-MS model for solving MS problems will be presented in the following papers: 1) a simple, fast and reliable method to determine MS in a given direction of power change is proposed; a technique that considers an error in a powers change forecast is also elaborated; 2). the closest MS models in $p$-norms are proposed and analyzed, their usage adequacy for a steady-state stability reserve assessment and for moving a system state into a power flow feasibility region with a minimum number and the volume of control actions is determined.

APPENDIX

To determine the interrelation between Lagrange multipliers, it is convenient to make use of the system power balance equation

$$\Delta P^b = P_b + \sum_{k \neq b} P_k - \pi = 0, \quad (A1)$$

where $\pi$ is system active power loss.

If to take advantage of the reduced gradient approach, i.e. use the variables set as in the load flow model, then differentiation of (A1) and (13) with respect to independent variables yield

$$\begin{aligned} \left[\partial \Delta P^b / \partial P\right] &= \left[\partial P_b / \partial P\right] + e^T - \left[\partial \pi / \partial P\right] = 0^T; \\ \left[\partial \Delta P^b / \partial Q\right] &= \left[\partial P_b / \partial Q\right] + e^T - \left[\partial \pi / \partial Q\right] = 0^T; \end{aligned} \quad (A2)$$

$$\begin{aligned} \left[\partial \Delta P_b / \partial P\right] &= \left[\partial P_b / \partial P\right] + \left[\partial \Delta P_b / \partial \delta\right]\left[\partial \delta / \partial P\right] + \\ &\quad + \left[\partial \Delta P_b / \partial V\right]\left[\partial V / \partial P\right] = 0^T; \\ \left[\partial \Delta P_b / \partial Q\right] &= \left[\partial P_b / \partial Q\right] + \left[\partial \Delta P_b / \partial \delta\right]\left[\partial \delta / \partial Q\right] + \\ &\quad + \left[\partial \Delta P_b / \partial V\right]\left[\partial V / \partial Q\right] = 0^T; \end{aligned} \quad (A3)$$

$$\begin{vmatrix} \partial \Delta P / \partial \delta & \partial \Delta P / \partial V \\ \partial \Delta Q / \partial \delta & \partial \Delta Q / \partial V \end{vmatrix} \begin{bmatrix} \partial \delta / \partial P & \partial \delta / \partial Q \\ \partial V / \partial P & \partial V / \partial Q \end{bmatrix} + [E] = [0]. \quad (A4)$$

Use of (A4) in (A3) yields

$$\begin{vmatrix} \partial \Delta P / \partial \delta & \partial \Delta P / \partial V \\ \partial \Delta Q / \partial \delta & \partial \Delta Q / \partial V \end{vmatrix}^T \begin{bmatrix} \partial P_b / \partial P^T \\ \partial P_b / \partial Q^T \end{bmatrix} = \begin{bmatrix} \partial P_b / \partial \delta^T \\ \partial P_b / \partial V^T \end{bmatrix}. \quad (A5)$$

Comparison (18) with (A5) and use of (A2) yield

$$\begin{aligned} \lambda_m^P &= \left(1 - \partial \pi / \partial P_m\right) \lambda_b^P; \\ \lambda_m^Q &= -\partial \pi / \partial Q_m \, \lambda_b^P \end{aligned}$$